\documentclass[12pt]{article}
\usepackage{geometry}
\usepackage{amsthm}
\usepackage{amsmath}
\usepackage{amssymb}
\usepackage[english]{babel}
\usepackage{caption}
\usepackage{float}
\usepackage{graphicx}
\usepackage{xfrac}
\usepackage{threeparttable}
\usepackage{subfig}
\usepackage{rotating}
\usepackage{hyperref}

\usepackage{stmaryrd}
\expandafter\def\csname opt@stmaryrd.sty\endcsname
{only,shortleftarrow,shortrightarrow}
\usepackage{extpfeil}
\usepackage{tcolorbox}

\usepackage{xcolor}
\usepackage{textgreek}
\usepackage{upgreek}

\title{An age-structured model of hepatitis B viral infection highlights the potential of different therapeutic strategies}

\author{Farzad Fatehi$^{1,2}$, Richard J. Bingham$^{1,2,3}$, Eric C. Dykeman$^{1,2}$, \\
Peter G. Stockley$^{4}$, Reidun Twarock$^{1,2,3,*}$}
\date{\footnotesize $^{1}$ York Cross-disciplinary Centre for Systems Analysis, University of York, York YO10 5GE, UK\\
$^{2}$ Department of Mathematics, University of York, York YO10 5DD, UK\\
$^{3}$ Department of Biology, University of York, York YO10 5NG, UK\\
$^{4}$ Astbury Centre for Structural Molecular Biology, University of Leeds, Leeds LS2 9JT UK\\
$^*$corresponding author: rt507@york.ac.uk \vspace{-0.6cm}}

\begin{document}

\maketitle

\begin{abstract}
Hepatitis B virus is a global health threat, and its elimination by 2030 has been prioritised by the World Health Organisation. Here we present an age-structured model for the immune response to an HBV infection, which takes into account contributions from both cell-mediated and humoral immunity. The model has been validated using published patient data recorded during acute infection. It has  been adapted to the scenarios of chronic infection, clearance of infection, and flare-ups via variation of the immune response parameters. The impacts of immune response exhaustion and non-infectious subviral particles on the immune response dynamics are analysed. A comparison of different treatment options in the context of this model reveals that drugs targeting aspects of the viral life cycle are more effective than exhaustion therapy, a form of therapy mitigating immune response exhaustion. Our results suggest that  antiviral treatment is best started when viral load is declining rather  than in a flare-up. The model suggests that a fast antibody production rate always lead to viral clearance, highlighting the promise of antibody therapies currently in clinical trials.
\end{abstract}

\section{Introduction}

Five viruses are known (HepA-E), from differing viral families, that infect the human liver resulting in inflammation, cirrhosis and liver cancer \cite{Zuckerman1996}. Collectively these impose a huge annual burden of death and suffering worldwide. Fortunately, progress is being made towards their treatment and prevention. HepC has been cured ($>$90\%) by directly-acting antiviral (DAA) treatment, principally via protease inhibitors, although there are still financial issues preventing their worldwide application. Despite pioneering use of recombinant DNA technology to create a safe and cheap prophylactic vaccine against Hepatitis B virus (HBV) \cite{Mackay1981,Murray1989} it is not universally deployed allowing the virus to infect $\sim$30 million people  each year.  These add to the very significant burden of the $\sim$292 million chronically infected individuals, that are the consequence of the failure of their immune systems to clear a primary infection \cite{World2017}. Treatment of these patients relies on generic replicase inhibitors that rapidly elicit resistance and interferon to boost antiviral responses. These treatments are largely ineffective leading to  $\sim$884,000 additional HBV-related deaths annually, the largest cause of liver cancer worldwide \cite{World2017}. HBV infected cells produce a significant excess of non-infectious particles, such as subviral Dane particles (SVPs) that occur at $\sim$1,000–100,000-fold higher concentration than the infectious virion, allowing them to  act as immune system decoys by sequestering antiviral antibodies \cite{Hu2017,Ciupe2014}.

HBV infected individuals that mount an adequate immune response tend to clear the infection. During the early stages of the infection the first contribution of the immune response comes from innate immunity which reduces viral spread and facilitates an adaptive immune response. Adaptive immunity is made of two components, {\it cell-mediated immunity} (CD4$^+$ and CD8$^+$ T cells) and {\it humoral immunity} (antibodies). CD4$^+$ T cells, also known as helper T cells, assist the activity of other immune cells by releasing cytokines. CD8$^+$ T cells are not only responsible for killing of infected cells but also induce the noncytolytic ``cure'' of such cells while antibodies neutralise virus particles and prevent infection of cells \cite{Guidotti1999,Abbas2014}.

Mathematical modelling provides new insights into the various aspects of viral infections and the impact of the immune response on their clearance. Ciupe et al. presented a model to study the role of cytolytic and noncytolitic immune responses and the time lag associated with effector cell activation and expansion during an acute HBV infection. It is hypothesised that cured cells and their progeny are less likely to get infected, preventing reinfection \cite{Ciupe2007,Ciupe2007b}. Later, Ciupe et al. looked into the dynamics of antibodies and showed that still having a strong cell-mediated immune response is crucial for the control of early infection in unvaccinated individuals \cite{Ciupe2014}. Fatehi et al. developed a mathematical model to take into account contributions from innate and adaptive immune responses, as well as cytokines. The model investigates the role of different parts of the immune response on viral dynamics \cite{Fatehi2018}. These models are based on systems of ordinary differential equations (ODE) or delay differential equations (DDE). Nelson et al. presented an age-structured model of human immunodeficiency virus (HIV) infection to study the impact of variations in the virion production rate and the death rate of infected cells over the course of the infection \cite{Nelson2004}. Although age-structured models are more complicated, they can provide more realistic dynamics \cite{Browne2016}.

Experimental studies have shown that persistent stimulation of effector cells may result in immune impairment, e.g., immune exhaustion \cite{Moskophidis1993,Zajac1998,Wherry2004}. In HBV infections, persistent antigen presentation by infected cells and exposure to high antigen loads plays an important role in CD8$^+$ T cell exhaustion \cite{Revill2019}. In order to analyse the impact of T cells exhaustion on viral dynamics, Johnson et al. introduced a variable that captures  the antigenic stimulus, called the level of exhaustion \cite{Johnson2011}. They assumed that T cells are inactivated dependent on the level of exhaustion and modelled it as a Hill function with a half-maximal constant called the exhaustion threshold. They showed that the exhaustion threshold has a significant impact on the ability of the immune response to control an infection \cite{Johnson2011}. Later, Conway and Perelson included T cell exhaustion into an HIV infection model and showed that the strength of cytotoxic T lymphocytes in killing productively infected cells, and the level of latently infected cells, determine the post-treatment outcome of the infection \cite{Conway2015}.

We recently introduced a model of intracellular HBV infection dynamics and used it for comparative analysis of different therapeutic strategies \cite{Fatehi2021}. The model reveals a two-phase behaviour in the release of non-infectious SVPs. Shortly after infection, a cell starts secreting SVPs. When the first intact virions are released, after $\sim$90 hours post-infection, SVP secretion stalls until the onset of a second secretion phase. In order to analyse the impacts of this behavior on the immune response dynamics, and in particular their interaction with anti-HBsAg antibodies, we have developed an age-structured model using infection kinetic parameters derived from our intracellular model. This model has been parameterised such that its outputs fit  data recorded from patients undergoing  an acute infection. We have  adapted to the scenarios of a chronic infection, immune clearance of the infection, and infection flare-ups via variation of the immune response parameters. The role(s) of cell-mediated and humoral immunity in an acute HBV infection, including the effects of immune response exhaustion, have been analysed. The impacts of DAA treatments targeting various steps of the virus life cycle are compared to blocking T cells exhaustion, which is known as exhaustion therapy. Optimal treatment regimes, such as the most effective treatment start times for the reduction of viral load, have been identified. Our analysis highlights the potential benefit  of antibody therapies currently in clinical trials, and allows modelling of the impacts of DAAs under development.

\section{Results}

\subsection{Model derivation}

In order to analyse HBV infection dynamics highlighting the roles of the immune response, we developed a detailed intercellular age-structured model of HBV infection. This model includes  uninfected target cells, $T(t)$, which are assumed to be created at rate $\lambda$ and to die at a rate $d$ \cite{Guedj2013,Conway2015}. Infected cells are structured by the age, $a$, of infection, $I(a,t)$ \cite{Nelson2004}. An infected cell produces complete virions, which are infectious, and incomplete particles which occur in three distinct forms: RNA containing particles; empty virions; and subviral particles (SVPs) which are either filamentous or spherical \cite{Hu2017}. The release dynamics of these particles from an infected cells has been studied precisely \cite{Fatehi2021}. The variables $V_c(t)$ indicates the number of complete virions which infect target cells at a rate $\beta$. All types of incomplete particles are non-infectious and are covered with HBV surface antigen (HBsAg), which enables them to act as decoys for the immune response by consuming HBsAg-specific antibodies, $A(t)$. We therefore just include the variable $V_i(t)$ into the model, which represents the total number of incomplete particles \cite{Hu2017}. The production rate of these particles depends on the age of an infected cell. The functions $P_c(a)$ and $P_i(a)$ show the production profiles of complete and incomplete particles, respectively, from infected cells of age $a$ \cite{Nelson2004}. We use the functions that are presented in Fig. 2c of \cite{Fatehi2021} as the base functions $P_c(a)$ and $P_i(a)$ (Figure S1). Since it has been shown that changing the intracellular model parameters will change the total number of released particles, we scale functions $P_c(a)$ and $P_i(a)$ with $\rho_1$ and $\rho_2$, respectively. We assume that $\rho_1$ and $\rho_2$ are the changeable parameters and fit them to patient data. Complete and incomplete particles are cleared, based on their half-life, at rates $d_c$ and $d_i$. HBsAg-specific antibodies, which are a component of the adaptive immune response called {\it humoral immunity}, are produced at rate $p_A$ proportional to antigen load, and are degraded at rate $d_A$. We also add a logistic growth to the antibody equation, which is maintained through homeostatic proliferation of memory B cells after infection. Antibodies can bind to complete and incomplete particles at rate $k_f$ to create complete virion-antibody complexes, $X_c$, and incomplete particle-antibody complexes, $X_i$, respectively. These complexes disassociate at rate $k_b$, or degrade at rate $d_x$ \cite{Ciupe2014}. The other component of the adaptive immune response is {\it cell-mediated immunity}, which occurs via the action of CD8$^+$ T cells, also referred to as effector cells, $E(t)$. These cells kill or cure infected cells. Since it has been hypothesised that cured cells lose their resistance to productive infection at a slow rate (in the order of $10^{-5}$ per day) \cite{Ciupe2007}, we model these two impacts in one reaction, where effector cells remove infected cells at rate $\mu$. Moreover, we assume infected cells die at rate $\delta$ due to infection or innate immune response, which is not included in the model directly \cite{Ciupe2014,Fatehi2018}. Effector cells are assumed to be produced at rate $\lambda_E$ and removed at rate $d_E$ in the absence of infection \cite{Ciupe2007}. During infection, proliferation of effector cells happens in an infected cell density dependent manner with a time delay, i.e. the density of antigen, where maximum proliferation rate is $\alpha$, and $\phi$ is the level of infected cells at which proliferation is half-maximal \cite{Johnson2011,Ciupe2007,Fatehi2019}. It has been argued that persistent exposure to high antigen loads is associated with CD8$^+$ T cell exhaustion \cite{Revill2019}. Therefore, functional effector cells are lost at maximal rate $\xi$, depending on the level of exhaustion, $Q$, which is implemented as a Hill function with coefficient $n$ and half-maximal constant $q_c$ \cite{Johnson2011}. The level of exhaustion, $Q$, is measured by integrating over the antigenic stimulus (${Y}/{(\phi+Y)}$) times the parameter $\kappa$ which is called ``blockade parameter'', and reduces exponentially with coefficient $d_q$, which is the rate of immune response recovery from exhaustion. The reduction of the blockade parameter is assumed to simulate the blockade of interaction between the inhibitory receptor programmed death 1 (PD-1) and its ligand (PD-L1) which restores CD8$^+$ T cell function \cite{Jagadish2015}.

\begin{figure}[H]
	\centering
	\includegraphics[width=0.6\linewidth]{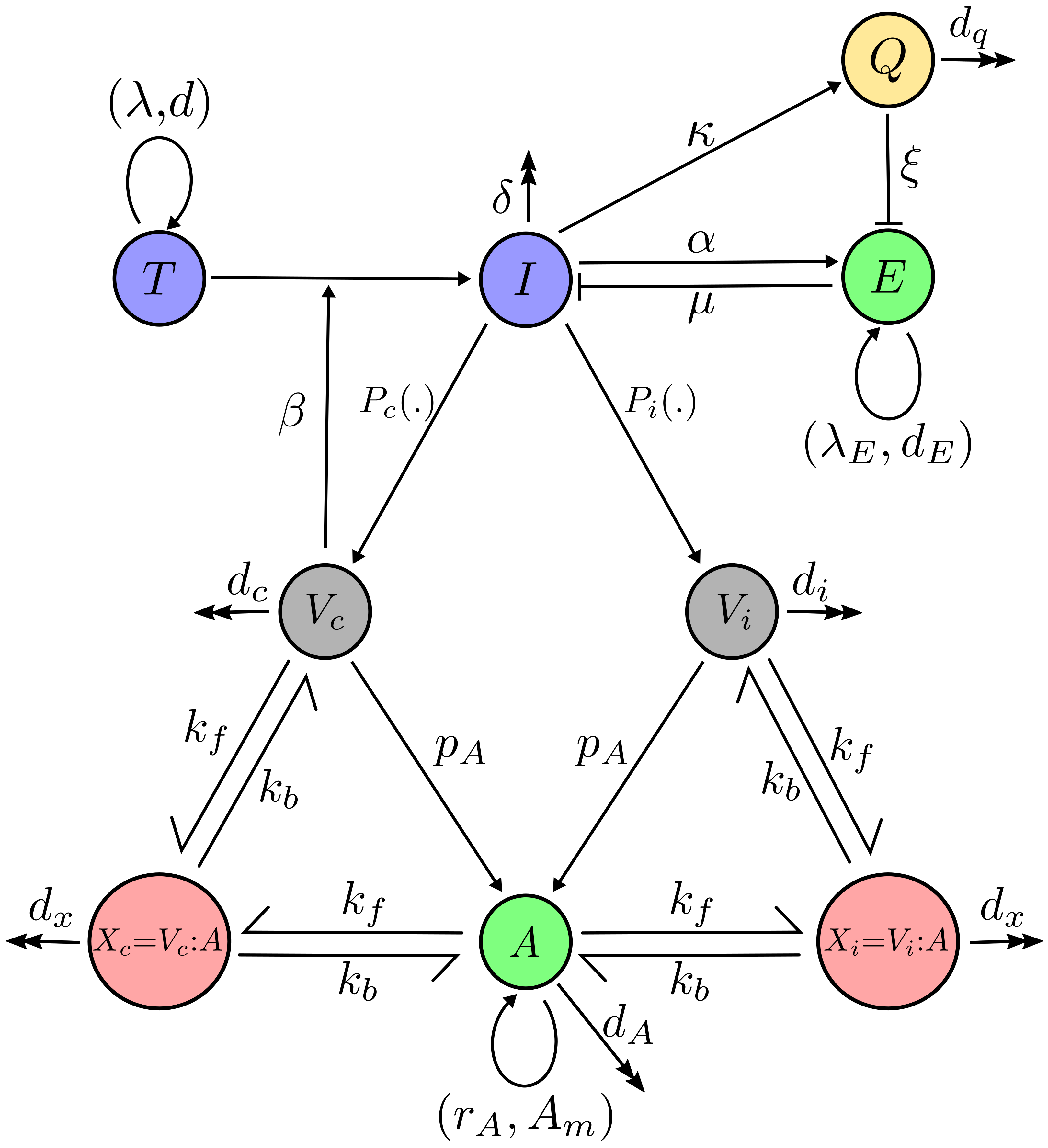}
	\caption{The components and allowed interactions of the model (\ref{model}). Purple circles show host cells (uninfected and infected cells), green circles indicate immune response (antibodies and effector cells), and gray circles show complete and incomplete particles. Pink circles represent complete virion-antibody and incomplete particle-antibody complexes. Yellow shows the level of exhaustion. Double arrow-headed lines show natural clearance. Bar-headed and single arrow-headed lines indicate destroying and production/proliferation of one cell population by/from another. Forward/backward arrows represent the binding/unbinding processes.}
	\label{diagram}
\end{figure}

With the above assumptions, the model, as illustrated in Fig. \ref{diagram}, takes on the form:
\begin{equation}\label{model}
\begin{array}{l}
    \displaystyle{\dfrac{dT}{dt}=\lambda - d T-\beta TV_c,}\\\\
    \displaystyle{\dfrac{\partial I(a,t)}{\partial t}+\dfrac{\partial I(a,t)}{\partial a}=-\delta I(a,t)-\mu I(a,t)E,}\\\\
    \displaystyle{\dfrac{dV_c}{dt}=\rho_1\int\limits_0^{\infty}P_{c}(a)I(a,t)da-d_{c}V_c-k_fAV_c+k_bX_{c},}\\\\
    \displaystyle{\dfrac{dV_i}{dt}=\rho_2\int\limits_0^{\infty}P_{i}(a)I(a,t)da-d_{i}V_i-k_fAV_i+k_bX_{i},}\\\\
    \displaystyle{\dfrac{dA}{dt}=p_A(V_c+V_i)+r_AA\left(1-\dfrac{A}{A_m}\right)-d_AA-k_fA(V_c+V_i)+k_b(X_c+X_i),}\\\\
    \displaystyle{\dfrac{dX_c}{dt}=k_fAV_c-k_bX_{c}-d_xX_c,}\\\\
    \displaystyle{\dfrac{dX_i}{dt}=k_fAV_i-k_bX_{i}-d_xX_i,}\\\\
    \displaystyle{\dfrac{dE}{dt}=\lambda_E+\alpha\dfrac{Y(t-\tau)}{\phi+Y(t-\tau)}E(t-\tau)-\xi\dfrac{Q^n}{{q_c}^n+Q^n}E-d_EE,}\\\\
    \displaystyle{\dfrac{dQ}{dt}=\kappa\dfrac{Y}{\phi+Y}-d_qQ,}
\end{array}
\end{equation}
where $\displaystyle{I(0,t)=\beta TV_c}$ and $\displaystyle{Y=\int\limits_0^{\infty}I(a,t)da}$.

\subsection{Within-host viral dynamics}

\begin{figure}[H]
	\centering
	\includegraphics[width=1\linewidth]{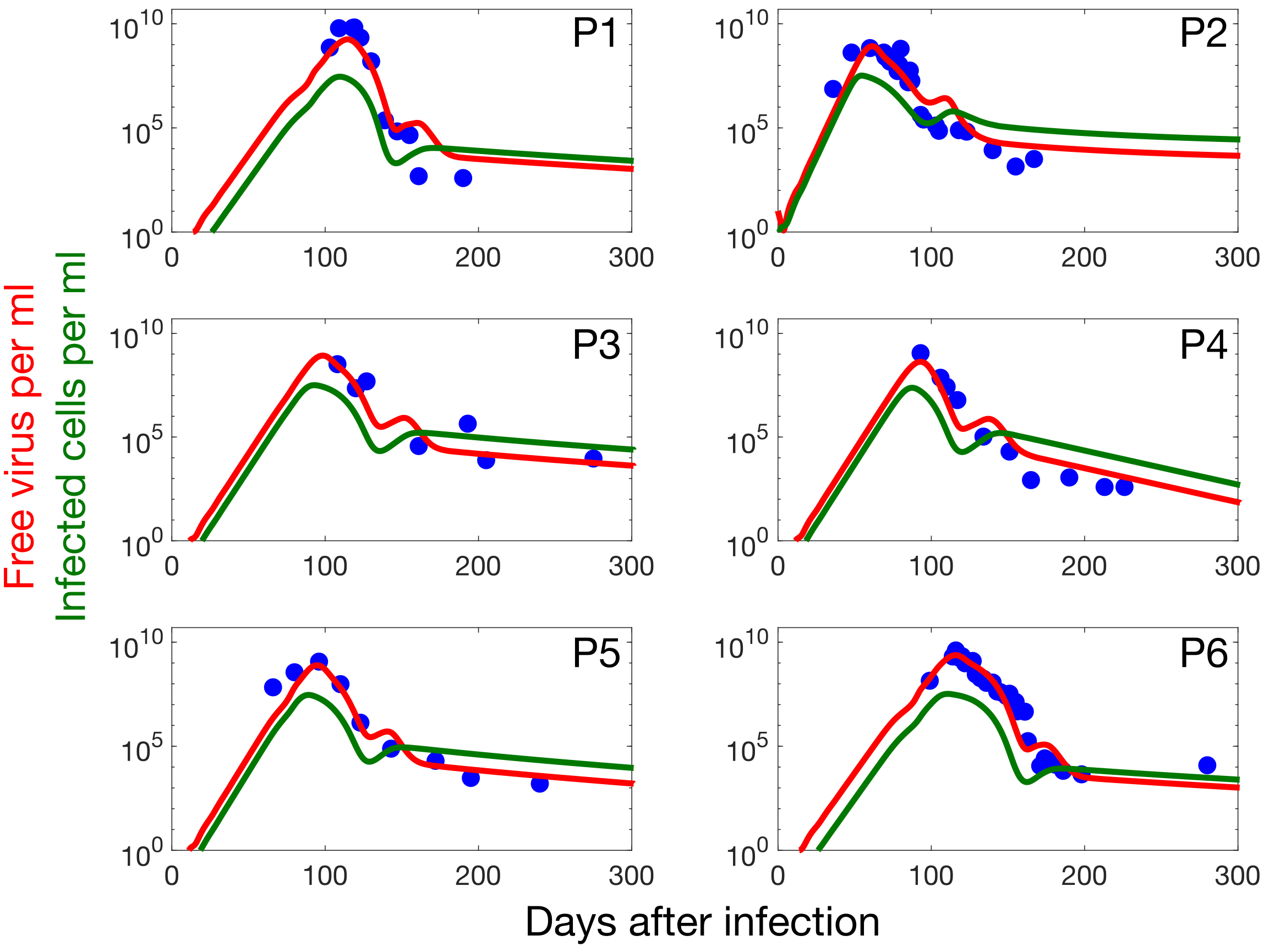}
	\caption{Paramaterising the model against patient data. The fitting of $V_c$ as given by model (\ref{model}) (red line) to patient data (\textcolor{blue}{$\bullet $}) indicates a biphasic decline in viral load and a slower decline in the level of infected cells compared with virions (green line).}
	\label{infected dynamics}
\end{figure}

The model (\ref{model}) was fitted to data of viral loads from 6 patients who were recorded during the acute stage of infection (Fig. \ref{infected dynamics}) \cite{Ciupe2014,Ciupe2007}. In this work, we also let the initial viral load ($V_c(0)$) be a variable. In patient number 2 (Fig. \ref{infected dynamics} P2) the peak in viral load occurs earlier than in other patients, which shows the initial viral load could be higher in this patient. We observe that for patient number 1, 3, 4, 5 and 6, $V_c(0)=0.33$ virion per ml as suggested by Ciupe et al. \cite{Ciupe2014}, but for patient number 2 $V_c(0)=10$ virion per ml. The parameter values derived from fitting of $V_c$ are presented in Table \ref{patient param}. Our model captures essential features of the viral load in all patients, including the first rapid decline in the level of viral load after the peak, followed by a slower decline, i.e. the biphasic viral decay from the peak \cite{Ciupe2007}. In all patients viral load eventually decreases to zero, matching the clinical outcomes in these patients with acute HBV infections.

The average fraction of infected cells at the peak of infection is 74.47\%, with a range of [60\%, 82\%] that is in good agreement with previously estimated values \cite{Ciupe2007,Ciupe2007b}, and the experimental observation that more than 75\% of hepatocytes were hepatitis B core protein antigen positive ($\mbox{HBcAg}^+$) in infected chimpanzees \cite{Guidotti1999}. Previous models suggested that the level of infected cells has a biphasic decline and decreases faster than the viral load \cite{Ciupe2007,Ciupe2007b}. However, it has been observed that the level of cccDNA (infected cells) has a slower decline compared with the level of free virus and remains detectable for more than 2 years after infection \cite{Wieland2004}. Our model predicts that infected cells decline slower compared with viral load (green lines Fig. \ref{infected dynamics}) and is in good agreement with Fig. 2 in Wieland et al. \cite{Wieland2004}.

\begin{sidewaystable}
\centering
\caption{Parameter best estimates}
\label{patient param}
\begin{tabular}{lccccccccccc}
\hline
Patient & $\beta\times10^{-9}$, day$^{-1}$ & $\mu\times10^{-4}$, day$^{-1}$ & $\delta$, day$^{-1}$ & $\rho_1$ & $\rho_2$ & $p_A\times10^{-4}$, day$^{-1}$ & $r_A$, day$^{-1}$ & $\alpha$ & $\tau$, day & $V_c(0)$, ml$^{-1}$ & RSS  \\\hline
1       & 0.35                                              & 1.00                                               & 0.0130                                 & 8.50      & 2.50      & 3.00                                               & 0.366                              & 2.20      & 6.50          & 0.33 & 0.29 \\
2       & 3.57                                              & 2.00                                               & 0.0210                                 & 3 .00       & 0.70      & 1.00                                               & 0.550                               & 1.50      & 4.50          & 10.00   & 0.14 \\
3       & 1.20                                               & 0.90                                             & 0.0170                                 & 3.20      & 0.70      & 1.00                                               & 0.391                              & 1.80      & 5.50          & 0.33 & 0.29 \\
4       & 1.70                                               & 2.00                                               & 0.0400                                  & 3.00        & 0.70      & 1.00                                               & 0.370                               & 1.80      & 5.00            & 0.33 & 0.24 \\
5       & 1.30                                               & 1.80                                             & 0.0180                                 & 3.50      & 0.70      & 1.00                                               & 0.430                               & 1.80      & 6.00            & 0.33 & 0.22 \\
6       & 0.35                                              & 1.00                                               & 0.0130                                 & 8.50      & 2.50      & 3.00                                               & 0.366                              & 2.20      & 8.70          & 0.33 & 0.13 \\\hline
median  & 1.25                                              & 1.40                                             & 0.0175                                & 3.35     & 0.70      & 1.00                                               & 0.381                             & 1.80      & 5.75         & 0.33 & -    \\
mean    & 1.41                                              & 1.45                                            & 0.0203                                  & 4.95     & 1.30      & 1.67                                            & 0.412                              & 1.88     & 6.03         & 1.94 & -    \\
std     & 1.19                                              & 0.54                                            & 0.0101                                  & 2.76     & 0.93     & 1.03                                            & 0.072                              & 0.27     & 1.49         & 3.95 & - \\\hline 
\end{tabular}
\end{sidewaystable}

\subsection{Adaptive immune response dynamics}

\begin{figure}[H]
	\centering
	\includegraphics[width=1\linewidth]{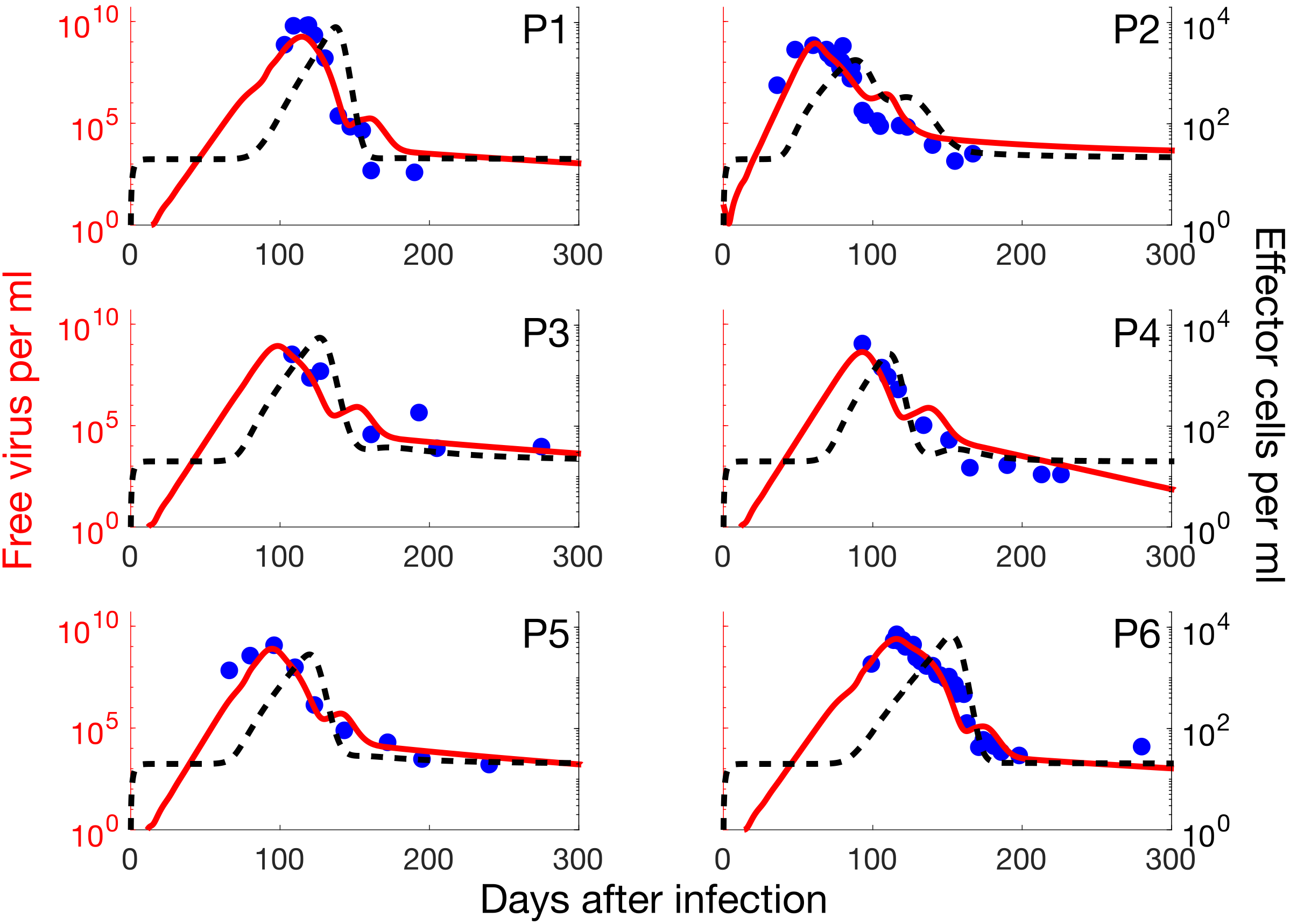}
	\caption{Time course of HBV infection. The maximal level of effector cells occurs 3-4 weeks after the peak in viral load. Red lines show the best fit to patient data (\textcolor{blue}{$\bullet $}). Black dashed lines show the levels  of effector cells predicted by the model (\ref{model}).}
	\label{viral dynamics}
\end{figure}

The level of serum alanine aminotransferase (ALT) is used as proxy for the dynamics of effector cells in HBV infection \cite{Ciupe2007,Ciupe2007b}. The peak in serum ALT occurs after the peak in viral load and the time lag between these two peaks is 3-4 weeks \cite{Webster2000,Ciupe2007}. Our model predicts that the average time lag in these 6 patients is 26.18 days (Fig. \ref{viral dynamics}) which is in a good agreement with previously reported values \cite{Webster2000,Ciupe2007}. To provide a more realistic dynamic of effector cells we included their exhaustion into our model. Figure S2 indicates that in each patient the onset of exhaustion coincides with a step increase in the level of effector cells. However, before the peak in exhaustion level, the immune response controls the infection, so that the level of exhaustion starts declining before the peak of effector cells is reached.

\begin{figure}[H]
	\centering
	\includegraphics[width=1\linewidth]{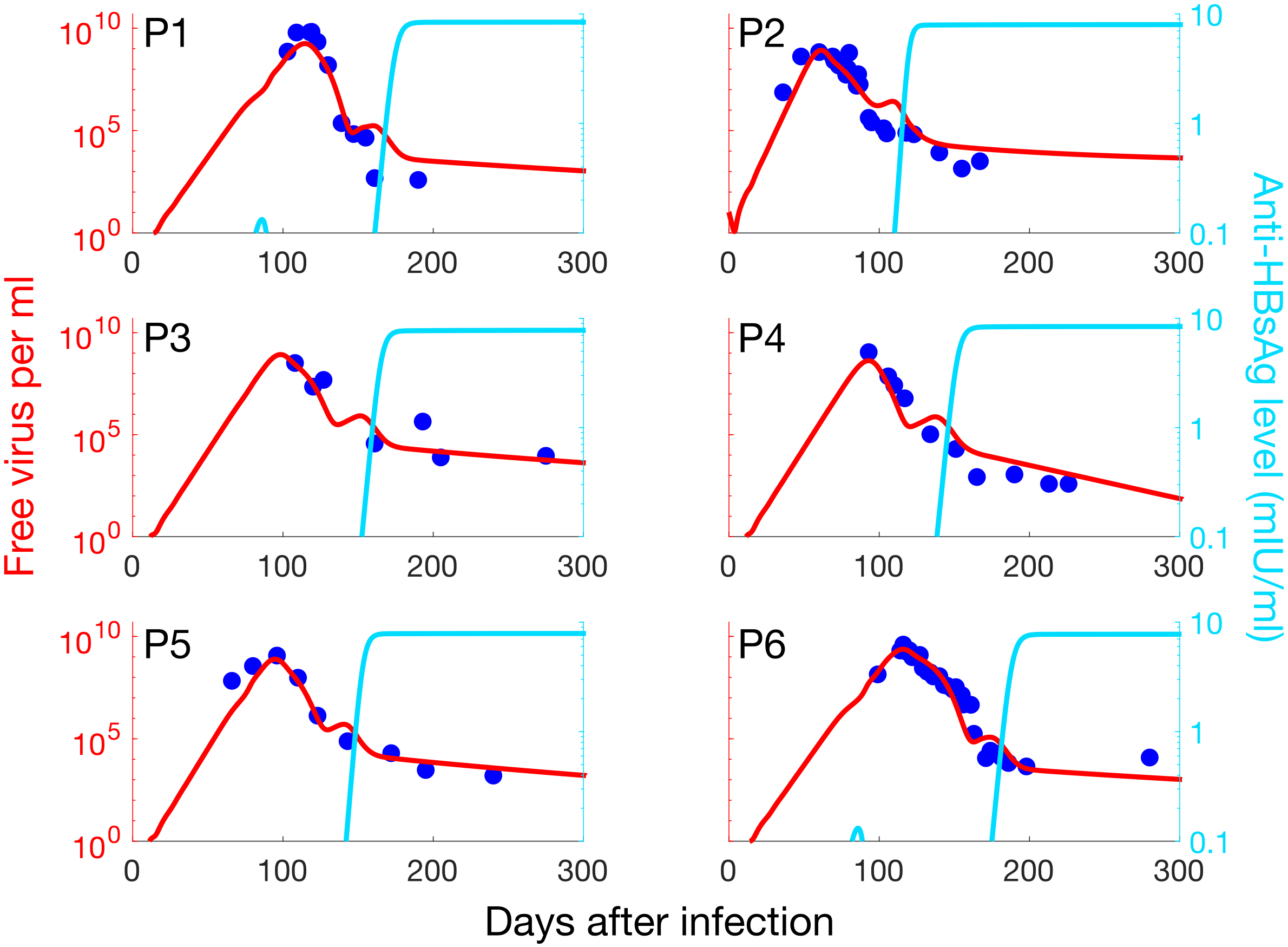}
	\caption{Time course of anti-HBsAg appearance. The red lines show the best fit to patient data (\textcolor{blue}{$\bullet $}). Cyan lines indicate the level of anti-HBsAg in mIU/ml unit.}
	\label{antibodies}
\end{figure}

Figure \ref{antibodies} indicates the level of antibodies against HBsAg (anti-HBsAg) in these patients. Anti-HBsAg levels $<10\mbox{ mIU/ml}$ ($8.5\times 10^{-4}\mbox{ mg/ml}$) are considered as negative \cite{Jack1999,Ciupe2014}. In unvaccinated patients, the level of antibodies is under 10 mIU/ml while they are still classed as infected, i.e. their HBV-DNA test is positive \cite{Guang2019}. Figure \ref{antibodies} shows that in all patients the level of antibodies is well below 10 mIU/ml during the peak in viral load and the first phase of viral decline. Later, when the level of virus has decreased significantly and effector cells are reduced to their basal level, it starts increasing but is still under 10 mIU/ml \cite{Guang2019}.

\begin{figure}[H]
	\centering
	\includegraphics[width=0.94\linewidth]{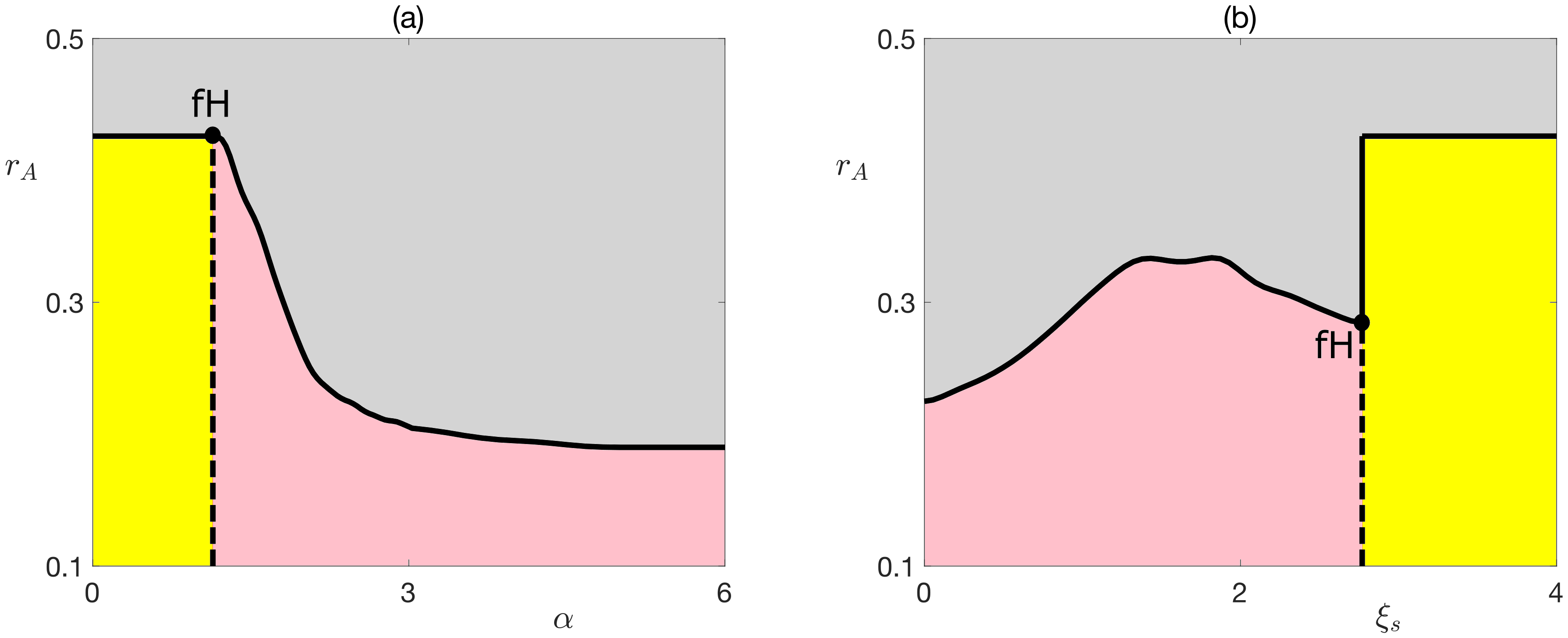}
	\caption{The stability of disease-free and chronic infection states. The parameters used are the median values from Table \ref{patient param}. Gray and yellow areas indicate the regions where the disease- free and chronic infection steady states are stable, respectively. Pink is the region where the system shows a stable periodic solution around a chronic infection steady state. Solid and dashed lines indicate the boundaries of the steady-state and Hopf bifurcation, respectively, and ``fH'' shows the location of the fold-Hopf bifurcation.}
	\label{sensitivity}
\end{figure}

\begin{figure}[H]
	\centering
	\includegraphics[width=0.94\linewidth]{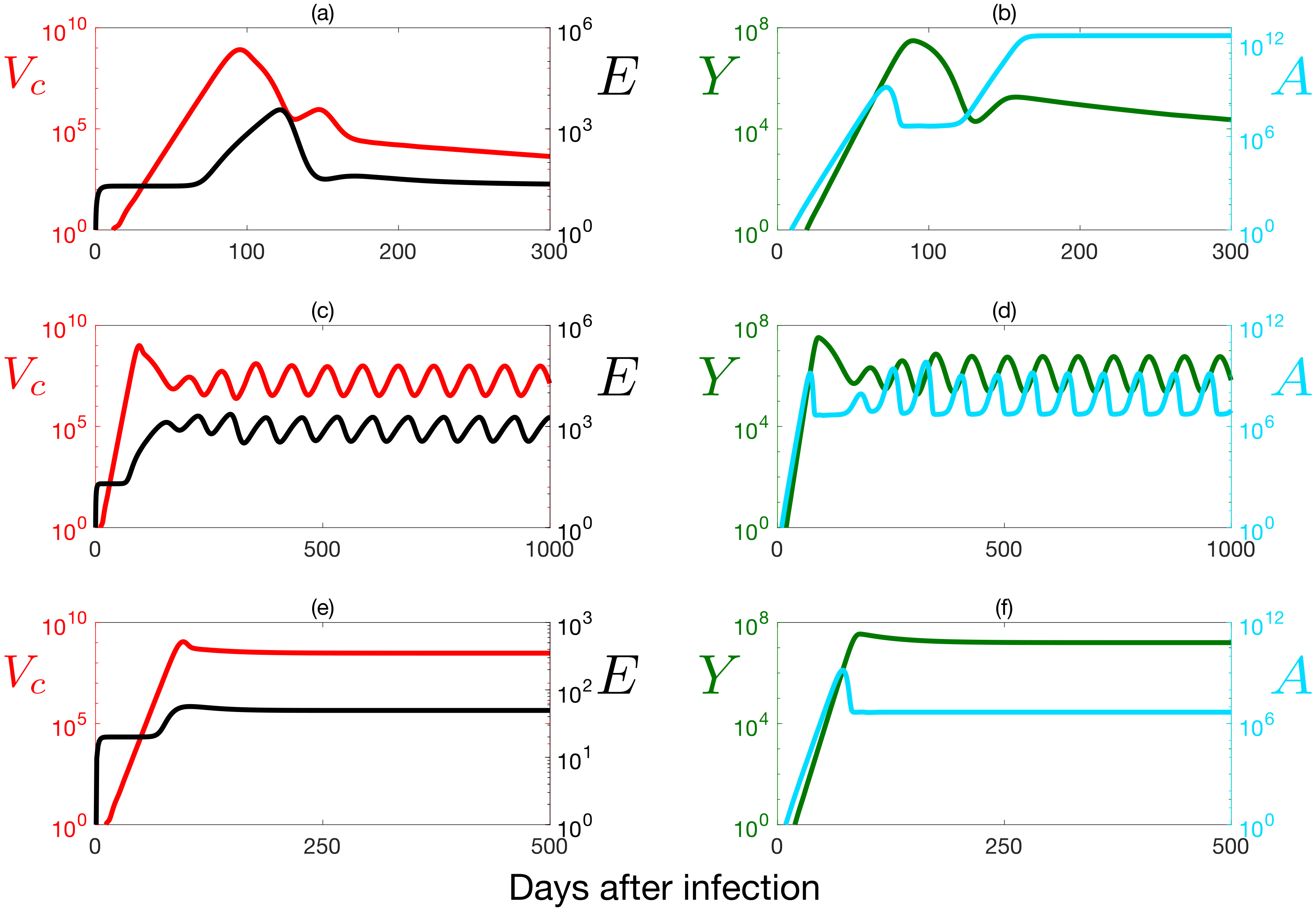}
	\caption{Examples of different types of infection dynamics: acute infections with immune clearance, and chronic infections with and without infection flare-ups. Numerical solutions of the model (\ref{model}) with parameter values from the median values of Table \ref{patient param}. (a) and (b) show acute infection. In (c) and (d) $\alpha$ (the maximum proliferation rate of effector cells) is reduced by 28\% ($\alpha=1.3$), indicating periodic oscillations around the chronic infection steady state (hepatitis flare). In (e) and (f) $\alpha$ is reduced by 56\% ($\alpha=0.8$), illustrating the case of a chronic infection.}
	\label{simulation}
\end{figure}

Now we study the impact of varying immune response parameters on viral dynamics. We focus on parameters $\alpha$ (growth rate of effector cells) and $\xi$ (the rate at which functional effector cells are lost due to the exhaustion) in the cell-mediated immune response which play crucial roles in the control of the infection \cite{Webster2000,Fatehi2018}. The other important parameter is the antibody proliferation rate ($r_A$) \cite{Ciupe2014}. Figure \ref{sensitivity} shows the stability and instability regions of the disease-free and chronic infection states as a function of these parameters. This figure shows that an increase in $r_A$ always stabilises the disease-free state. However, when $r_A$ is low ($<0.22$), increasing $\alpha$ or decreasing $\xi$ cannot stabilise the disease-free state, However, it destabilises the chronic steady state due to a Hopf bifurcation, so one observes stable oscillations (Fig. \ref{simulation}c and d). Biologically, this behaviour is called ``flare-ups'', where through the interactions between viral infection and immune response we observe a periodic behaviour \cite{Chang2014,Perrillo2001,Fatehi2018}. 

\subsection{Effects of current clinical therapies}

There are currently two approved anti-HBV therapies for clinical use. Interferon (IFN)-based therapy and nucleot(s)ide analogues (NAs) \cite{Lok2017}, but more therapeutic strategies will be required to meet the global WHO challenge to make HBV infection a treatable condition by 2030 \cite{Revill2019}. We recently compared current DAAs with  several possible future treatment options using our intracellular model \cite{Fatehi2021}. The latter include capsid assembly modulators (CAMs), especially based on heteroaryldihydropyrimidines HAPs \cite{Stray2005}. Recently we showed that both HCV and HBV regulate the assembly their nucleocapsids around ssRNA forms of their genomes (e.g. the pgRNA of HBV) at least in part via sequence-specific core protein-genomic RNA interactions at multiple sites, termed Packaging Signals (PSs) \cite{Stewart2016,Patel2017,Twarock2019}.  CAMs are at an advanced stage of development \cite{Zoulim2017,Lahlali2018,Zoulim2020}, whilst the vital roles of PS-core interactions in HBV nucleocapsid assembly, and their targeting by small ligands are at a much earlier stage.The comparative analysis reveals strong potential synergistic effects between them. We showed that starting these treatment options at different times post infection reduces the viral production rate by a factor $(1-\epsilon_1)$, where $0\leq\epsilon_1\leq 1$ is the drug efficacy \cite{Fatehi2021}. IFN-based therapy and CAMs also block {\it de novo} infections \cite{Guo2017,Fatehi2018}. Mathematically, one can represent this effect by a modified transmission rate $(1-\epsilon_2)\beta$, where $0\leq\epsilon_2\leq 1$ is the drug efficacy \cite{Fatehi2018,Fatehi2021}. Here we refer to these types of treatments as antiviral therapy. The other suggested form of therapy is the recovery of the CD8$^+$ T cells from exhaustion (exhaustion therapy) \cite{Revill2019}. The impact of this treatment option can be modelled by a modified blockade parameter $(1-\eta)\kappa$, where $0\leq\eta\leq 1$ is the efficacy of the treatment \cite{Jagadish2015}. The new equations take the following form

\begin{equation}\label{model treatment}
\begin{array}{l}
    \displaystyle{\dfrac{dT}{dt}=\lambda - d T-(1-\epsilon_2)\beta TV_c,}\\\\
    \displaystyle{\dfrac{dV_c}{dt}=\rho_1(1-\epsilon_1)\int\limits_0^{\infty}P_{c}(a)I(a,t)da-d_{c}V_c-k_fAV_c+k_bX_{c},}\\\\
    \displaystyle{\dfrac{dQ}{dt}=\kappa(1-\eta)\dfrac{Y}{\phi+Y}-d_qQ.}
\end{array}
\end{equation}

The total drug effectiveness $\epsilon_{tot}$, which is defined as $\epsilon_{tot}=1-(1-\epsilon_1)(1-\epsilon_2)$ will allow us to determine a critical drug efficacy, $\epsilon_c$, where the infection gets cleared for $\epsilon_{tot}>\epsilon_c$ \cite{Dahari2009,Fatehi2018}.

\begin{figure}[H]
	\centering
	\includegraphics[width=0.95\linewidth]{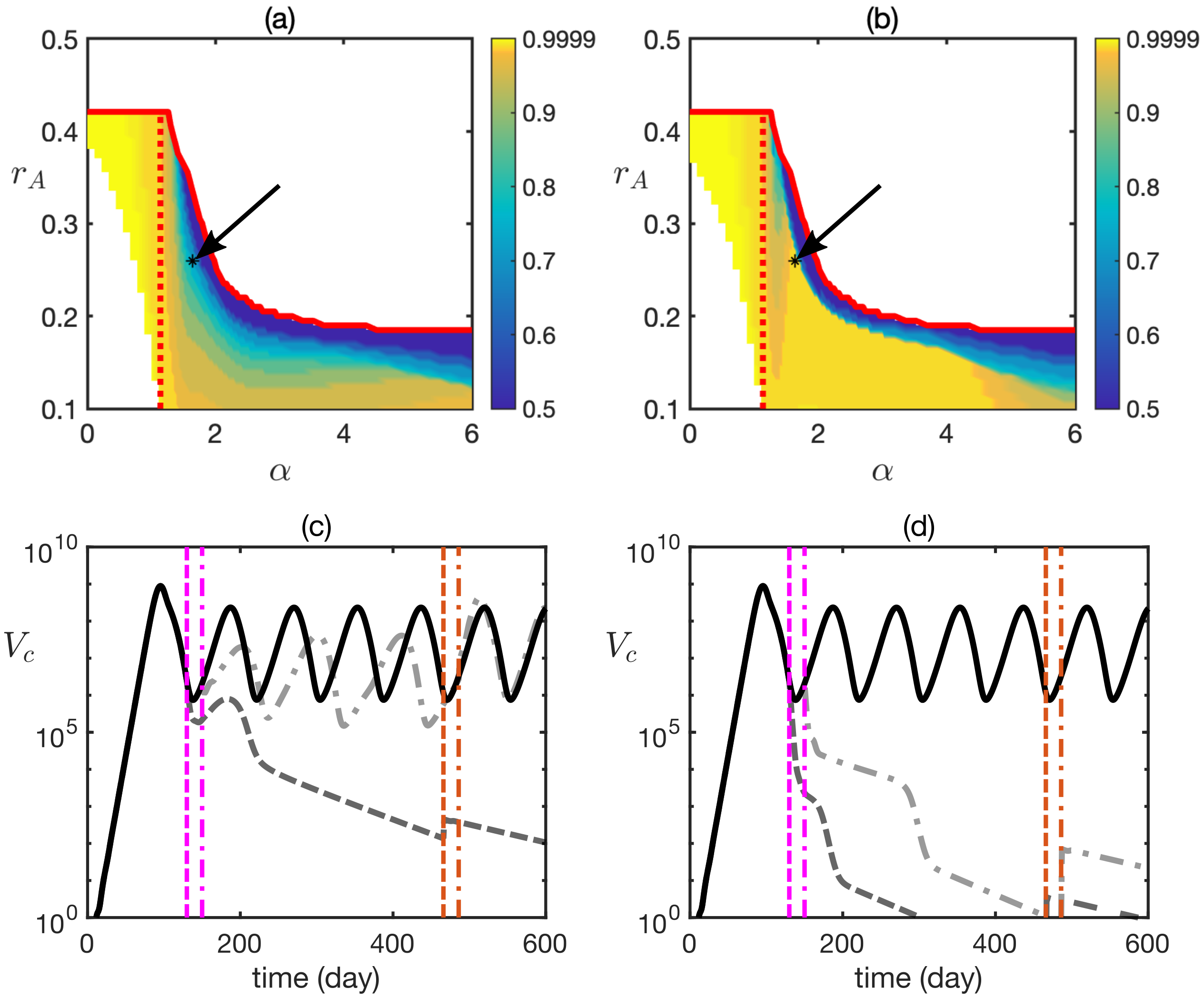}
	\caption{Start of antiviral therapy when viral load is in a declining phase is more effective. (a) and (b) show the minimal total efficacy ($\epsilon_{tot}$) that is required to clear the infection following a 48 weeks therapy starting at 130 and 150 days post infection (dpi), respectively. The white area indicates a stable disease-free state. The red solid curves indicate the onset of the steady-state bifurcation, whereas the red dotted lines of the Hopf bifurcation of the chronic state. The black hatched area indicates the region where an efficacy of $\le 99.99\%$ is ineffective. The black star (indicated with an arrow) shows the point at which the numerical simulations ((c) $\epsilon_{tot}=0.7$ and (d) $\epsilon_{tot}=0.99$) have been performed. Magenta and brown vertical lines indicate the start and end of treatment. The dashed and dash-dotted lines, represent the effects of treatments on viral load, for treatment starts 130 and 150 dpi, respectively.}
	\label{Treatment start}
\end{figure}

IFN-based therapy, which results in higher rates of hepatitis B e antigen (HBeAg) and HBsAg loss compared with NA therapy, is usually administered for around 48 weeks \cite{Lok2017}. Thus, 48 weeks after the start of treatment we remove its effect on the model to determine the efficacy required to avoid a viral rebound after the removal of treatment. This will help us to find an optimal efficacy that a new therapy should have to be effective, thus addressing an urgent need identified in \cite{Lok2017}. Figure \ref{Treatment start} shows the impact of starting antiviral therapy at different times post infection. In Fig. \ref{Treatment start}a treatment start is 130 dpi, while in Fig. \ref{Treatment start}b it is 150 dpi. These figures indicate that in a region where the system shows a periodic behaviour, an earlier treatment start during the declining phase of the viral dynamics, is more effective (Figs. \ref{Treatment start}c and d). Video S1 shows the minimal total efficacy ($\epsilon_{tot}$) that is required to clear the infection following a 48 weeks therapy starting at various times between 50 dpi to 160 dpi. It indicates that treatment start at 50-80 dpi (around a month before the first viral peak) is most effective. 

\noindent
It also shows that treatment start at 90-120 dpi (close to the first viral peak) is not that effective, in particular when the chronic steady state is stable. After the first viral peak, if the chronic steady state is stable, the treatment start time is not important as the infection is in a stable state. However, in the case of a periodic solution, starting treatment when the viral load is declining requires a lower efficacy treatment to be successful compared with when the viral load is in an increasing phase. These results indicate that a decline in the level of free virus in a patient, can be a sign of an acute infection, or correspond to the declining phase before a flare-up. Therefore, it is crucial to start the treatment, rather than wait to start it after the flare-up. Video S2 shows the minimal efficacy required to clear the infection following 48 weeks of exhaustion therapy starting at in between 50 dpi and 160 dpi. Surprisingly, it shows that this treatment is not as effective as antiviral therapy targeting the viral life cycle and the treatment start point has no significant impact on the outcome. This can be seen from the fact that this treatment option is only effective in a small area of the $r_A$-$\alpha$ parameter plane, corresponding to an area in which antiviral therapy targeting the viral life cycle is also effective even for low drug efficacy.

\section{Discussion}

In this paper we have developed an age-structured model for the immune response to HBV infections. It is based on the particle release profiles from a detailed intracellular model \cite{Fatehi2021}. Our age-structured model focuses on two components of the adaptive immune response known as humoral (antibodies) and cell-mediated (CD8$^+$ T cells) immunity. To provide more realistic viral dynamics and robust predictions, we fitted our model to patient data. Patient data indicate a biphasic decrease in the viral load which is captured well by the model. Previous models have predicted that the number of infected cells declines faster than viral load \cite{Ciupe2007,Ciupe2007b}, but it has been shown clinically that the level of cccDNA (infected cells) has a slower decay compared with the viral load \cite{Wieland2004}. Interestingly, our model reflects this trend. Regarding the dynamics of the immune response our model shows that the peak in the level of effector cells occurs around 26 days after the peak in viral load, which is in good agreement with previously reported values \cite{Webster2000,Ciupe2007}. The antibody level stays well under 10 mIU/ml during the peak and the first phase of viral decline. Later, it starts increasing, but it remains under 10 mIU/ml \cite{Guang2019}.

We also performed a stability analysis of the model to study the impact of varying parameter values on the outcome of the infection. The model suggests that increasing $r_A$ (the proliferation rate of antibodies) always stabilises the disease-free state. When $r_A$ is low, decreasing $\xi$ (the rate at which functional effector cells are lost due to exhaustion) or increasing $\alpha$ (expansion rate of effector cells) will not stabilise the disease-free state, but the chronic steady state loses its stability and one observes stable oscillations. Biologically, this behaviour is called ``flare-ups'' \cite{Chang2014,Perrillo2001,Fatehi2018}. Therefore, our model illustrates the importance of the {\it humoral immunity} in viral clearance.

Current therapeutic strategies either block the formation of new virions and viral entry (antiviral therapy) or recover the exhausted CD8$^+$ cells (exhaustion therapy). We used our model to compare the impacts of these different treatment options. Our results suggest that exhaustion therapy is not as effective as antiviral therapy, and that the start of treatment does not have a significant impact on the outcome of exhaustion therapy. However, treatment start plays a crucial role in the outcome of antiviral therapy. Our model shows that around one month before the peak in viral load would be the best time to start antiviral therapy, illustrating the importance of identifying cases early into an infection. Since this is not always practicable, we also analysed the impact of a treatment start later during infection. Whilst treatment start time is not significant in the context of stable chronic steady states, it matters for periodic solutions, where a treatment start during the  declining phase of viral load is more effective. As declining levels of free virus in a patient can indicate the declining phase before a flare-up, it is crucial to start treatment immediately rather than wait until after a flare-up. Our model moreover illustrates that antibodies play an important role in suppression of viral load, supporting strategies to use monoclonal antibodies as therapy against chronic Hepatitis B viral infections. Indeed, monoclonal anti-HBsAg antibody drugs are currently investigated in clinical trials \cite{Gao2017,Cerino2019,Alexopoulou2020}, and our model suggests that they should be highly efficient ways of combating chronic hepatitis B.

One critical hurdle to such developments has been the restriction of HBV infection to people and other higher primates, restricting the rapid development of new drugs in animal models. The development of humanised transgenic mice, available with and without a human immune system, will hopefully permit much faster experimental development \cite{Dusseaux2017,Li2018}. This model and others will be useful for the evaluation of any novel therapeutic strategies being developed.

\section{Materials and methods}

\subsection{Patient data}

Our study uses patient data comprising 6 patients whose their viral loads were recorded during the acute stage of infection and are reported in Ciupe et al. \cite{Ciupe2014,Ciupe2007}. The data are published in the Supporting Information of Ciupe et al. \cite{Ciupe2014}.

\subsection{Parameter estimation}

It has been estimated that the total number of liver cells in an adult is equal to $2\times 10^{11}$ and only 60\% of them are hepatocytes \cite{Ciupe2007,Michalopoulos2007,Kmiec2001,Goyal2017}. We assume that there are 3 liters of serum in an adult, containing complete and incomplete particles \cite{Murray2015}. Therefore, $\lambda/d=0.6\times 2\times 10^{11}/3000=4\times 10^7\mbox{ cells/ml}$. The death rate of target cells is estimated to be $0.01\mbox{ day}^{-1}$ \cite{Guedj2013,Kitagawa2018}, so $\lambda=4\times 10^5$. As in unvaccinated patients with positive HBV-DNA test the antibody level is under $10\mbox{ mIU/ml}=8.5\times 10^{-4}\mbox{ mg/ml}=3.4\times 10^{12}\mbox{ molecules/ml}$ \cite{Guang2019}, we assume that $A_m=3.4\times 10^{12}\mbox{ molecules/ml}$. For parameters $\phi$ (the level of infected cells at which growth is half-maximal) and $\xi$ (the rate at which functional effector cells are lost due to exhaustion) various fixed values have been determined previously \cite{Johnson2011,Conway2015}. We thus set $\phi=10^6 \mbox{ per ml}$ and $\xi=1\mbox{ day}^{-1}$. These values provide a good fit to the viral load data. All parameter values adapted from the literature are presented in Table \ref{fixed param} together with a pointer to the reference they have been adapted from. The remaining parameters are estimated by fitting the model (\ref{model}) to the viral load data using a modified version of DKLAG6 for the numerical simulations of the model (\ref{model}) (see \href{http://www.radford.edu/~thompson/ffddes/index.html}{www.radford.edu} for more detail)
\cite{Thompson2006} and an implementation of the Nelder-Mead algorithm (see \href{https://people.sc.fsu.edu/~jburkardt/f_src/asa047/asa047.html}{people.sc.fsu.edu} for more detail)
\cite{O1971}.

\begin{table}[H]
\centering
\caption{Parameter values adapted from the literature}
\label{fixed param}
\begin{tabular}{cll}
\hline
Parameter   & Value & Reference \\\hline
$\lambda$   & $4\times 10^5 \mbox{ per ml}$                             & \cite{Ciupe2007,Goyal2017,Murray2015}\\
$d$         & $0.01 \mbox{ day}^{-1}$                                   & \cite{Guedj2013,Kitagawa2018}\\
$d_c$       & 0.67 $\mbox{ day}^{-1}$                                   & \cite{Ciupe2014}\\
$d_i$       & 0.67 $\mbox{ day}^{-1}$                                   & \cite{Ciupe2014}\\
$k_f$       & $10^{-10} \mbox{ ml}\mbox{ molecule}^{-1}\mbox{day}^{-1}$ & \cite{Ciupe2014}\\
$k_b$       & 10 $\mbox{ day}^{-1}$                                     & \cite{Ciupe2014}\\
$A_m$       & $3.4\times 10^{12} \mbox{ per ml}$                        & \cite{Ciupe2014,Guang2019}\\
$d_A$       & 0.033 $\mbox{ day}^{-1}$                                  & \cite{Ciupe2014}\\
$d_x$       & 2.7 $\mbox{ day}^{-1}$                                    & \cite{Ciupe2014}\\
$\lambda_E$ & 10 $\mbox{ per ml}$                                       & \cite{Ciupe2007b}\\
$d_E$       & 0.5 $\mbox{ day}^{-1}$                                    & \cite{Ciupe2007b}\\
$q_c$       & 10                                                        & \cite{Johnson2011}\\
$n$         & 3                                                         & \cite{Johnson2011}\\
$\kappa$    & 1 $\mbox{ day}^{-1}$                                      & \cite{Johnson2011}\\
$d_q$       & 0.1 $\mbox{ day}^{-1}$                                    & \cite{Johnson2011}\\\hline
\end{tabular}
\end{table}

\section*{Acknowledgements}
RT acknowledges funding via an EPSRC Established Career Fellowship (EP/R023204/1) and a Royal Society Wolfson Fellowship (RSWF/R1/180009). RT \& PGS acknowledge support from a Joint Wellcome Trust Investigator Award (110145 \& 110146), and funding via the UK MRC (MR/N021517/1).

\section*{Supplementary Material}

\begin{figure}[H]
	\centering
	\includegraphics[width=\linewidth]{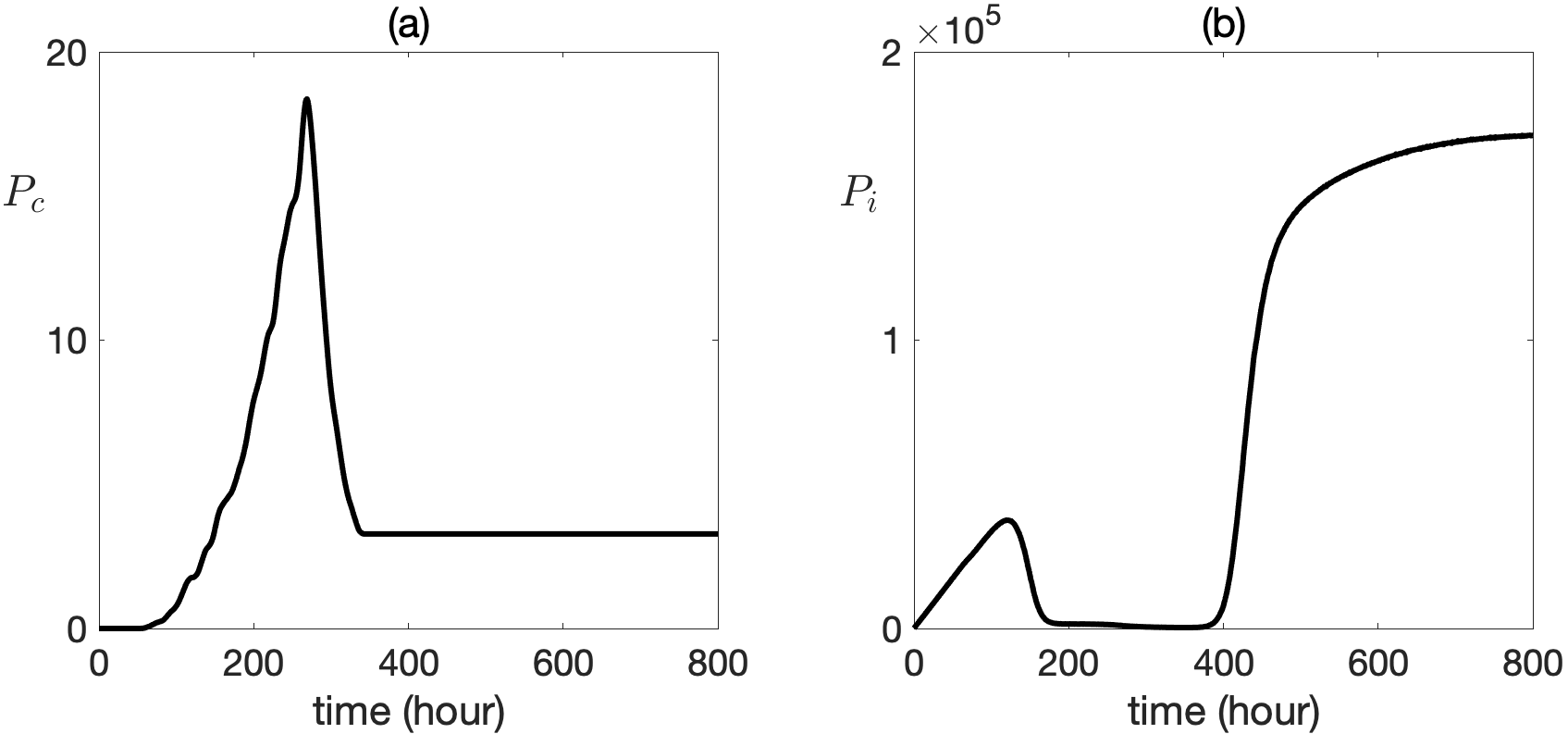}
	\caption*{ Figure S1: The production profiles of particles. (a) and (b) indicate the release profiles of complete ($P_c(.)$) and incomplete ($P_i(.)$) particles, respectively.}
	\label{figs1}
\end{figure}

\begin{figure}[H]
	\centering
	\includegraphics[width=\linewidth]{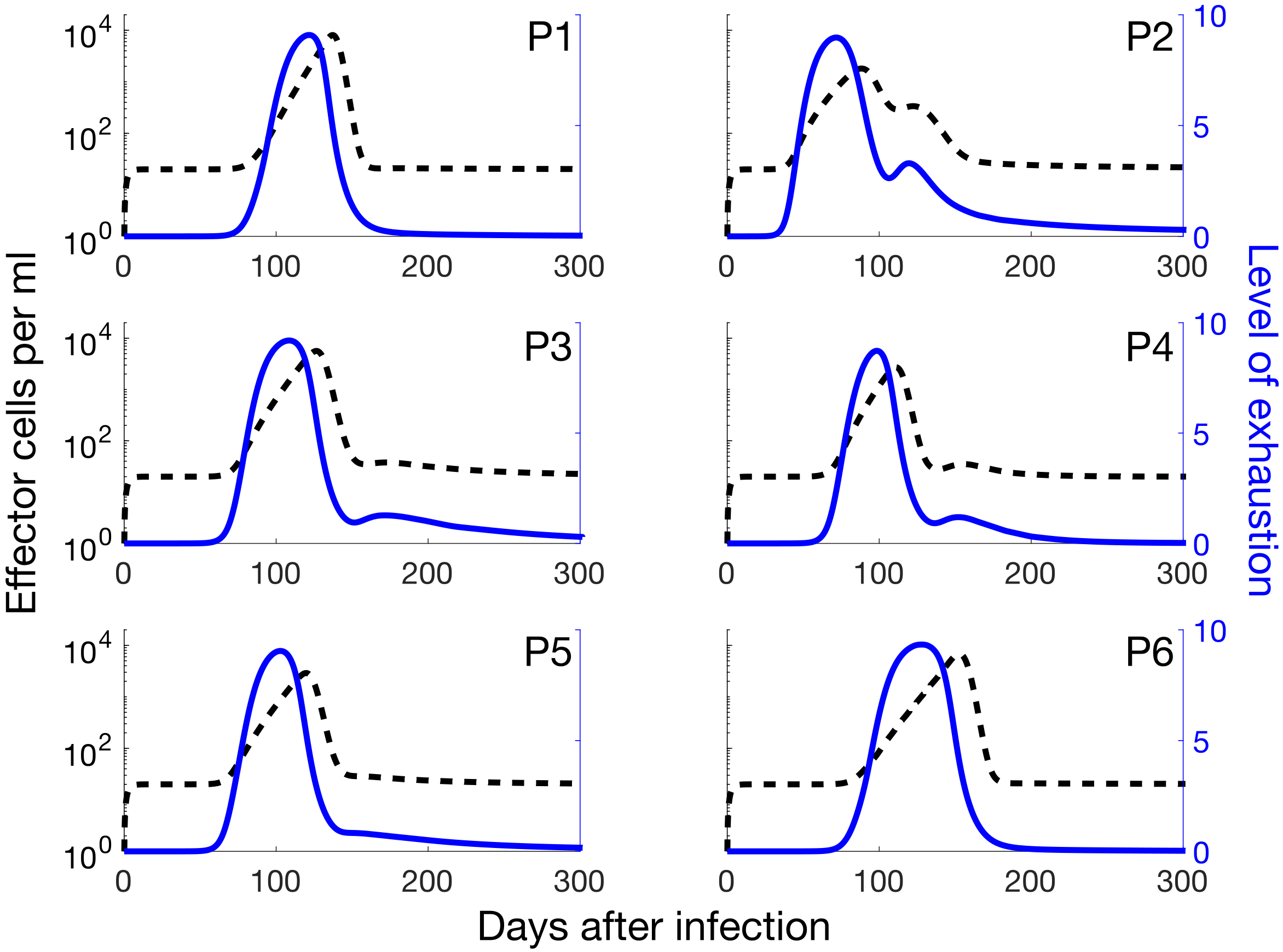}
	\caption*{Figure S2: The level of T cells exhaustion declines before the peak of effector cells in acute HBV infections. Blue lines indicate the level of exhaustion and black dashed lines show the dynamics of effector cells.}
	\label{figs2}
\end{figure}

\noindent
Video S1: The minimal total efficacy ($\epsilon_{tot}$) that is required for an antiviral therapy to clear the infection following a 48 weeks therapy starting at various times between 50 dpi and 160 dpi. The white area indicates a stable disease-free state. The red solid curves indicate the onset of the steady-state bifurcation, whereas the red dotted lines of the Hopf bifurcation of the chronic state. The black hatched area indicates the region where an efficacy of $\le 99.99\%$ is ineffective.

\noindent
Video S2: The minimal efficacy ($\eta$) that is required for exhaustion therapy to clear the infection following a 48 weeks therapy starting at various times between 50 dpi and 160 dpi. The white area indicates a stable disease-free state. The red solid curves indicate the onset of the steady-state bifurcation, whereas the red dotted lines of the Hopf bifurcation of the chronic state. The black hatched area indicates the region where an efficacy of $\le 99.99\%$ is ineffective.


\end{document}